# Exciting surface plasmons with transformation media

**Carlos García-Meca*, Rubén Ortuño, Javier Martí, and Alejandro Martínez**

*Nanophotonics Technology Center, Universidad Politécnica de Valencia, Camino de Vera s/n 46022, Valencia, Spain*

*E-mail: cargarm2@ntc.upv.es

**Abstract.** We present a way of exciting surface plasmon polaritons along non-patterned metallic surfaces by means of a flat squeezing slab designed with transformation optics. The slab changes the dispersion relation of incident light, enabling evanescent coupling to propagating surface plasmons. Unlike prism couplers, the proposed device does not introduce reflections at its input interface. Moreover, its compact geometry is suitable for integration. A feasible dielectric implementation of the coupler is suggested. Finally, we show that the angular response of the device can be engineered by using a non-uniform compression factor. As an example, we design a coupler with a half-power angular bandwidth 2.5 times higher than that of a conventional dielectric coupler.

---

## Introduction

Surface plasmon polaritons (SPPs) are hybrid electron-photon excitations that are trapped at the interface between a dielectric and a conductor [1-3]. This property of SPPs allows us to concentrate electromagnetic fields at the nanoscale by using subwavelength metallic structures. The branch of plasmonics exploits this unparalleled light-concentration ability of metals for a wide range of applications [2-5]. These include the miniaturization of photonic circuits, modulators and photodetectors, the enhancement of non-linear phenomena, the realization of extremely sensitive biosensors, and the improvement of the efficiency of photovoltaic cells. Recently, the use of transformation optics (TO) has been proposed to fully control the propagation of SPPs [6-9]. This technique enables us to engineer electromagnetic space by implementing arbitrary geometries and coordinate transformations with suitable media [10-12]. This way, TO makes it possible to design a variety of devices for SPPs such as cloaks, beam shifters, extreme bends, lenses, and wave adapters [6-9]. Although all these works have focused on flow control, SPPs need to be excited before they can be manipulated. Due to their bound nature, the momentum of SPPs is always higher than that of free-space photons of the same frequency [1-3]. Therefore, it is not possible to directly excite SPPs with free-space light. There exist several techniques to provide the required additional momentum for far-field excitation of SPPs. The most important ones are based on gratings or prism couplers [1-3,13]. The former consists of structuring periodically the metal surface or surrounding dielectric. If the gratings are sufficiently deep, the SPP dispersion relation (DR) can be significantly altered by the periodic pattern and photonic bandgaps and localized modes may appear [2,3]. The latter is based on the use of a prism placed next to a thin metal film. For instance, in the Kretschmann configuration, light entering the prism at a right angle increases its momentum by a factor equal to the prism index. When light reaches the prism-metal interface, total internal reflection occurs and the wave evanescently tunnels to the other metal side (in contact with air), where the SPP is excited [3,13]. The prism shape imposes a minimum size for the coupler that may not be suitable for integration. Here we propose an alternative way to excite SPPs with the help of TO. Specifically, we show that a properly-designed flat slab that performs a spatial compression can play the role of a prism coupler with important advantages.

## Theory

Light squeezers based on TO have been extensively studied from the spatial viewpoint [14-18]. However, although there exist some works that link Fourier optics and TO [19], little attention has been paid to the properties of squeezers in the Fourier domain. We begin by analyzing this kind of devices regarding them as $k$-space filters. For simplicity, we will focus on a two-dimensional problem (invariant in the $y$-direction), although the results could be extended to three dimensions. In flat-space TO, one starts from a virtual space and performs a certain coordinate transformation in such a way that electromagnetic fields are distorted in the desired manner. It is possible to implement this deformation in real physical space by filling it with the appropriate relative permittivity $\varepsilon^{ij}$ and permeability $\mu^{ij}$ [10,12]. Our compressing device will result from transforming a rectangular region A'B'C'D' in virtual empty space into region ABCD in physical space [Fig. 1(a-b)]. Both spaces are described by a Cartesian system (after reinterpretation) with coordinates $x'$, $y'$, $z'$ and $x$, $y$, $z$.

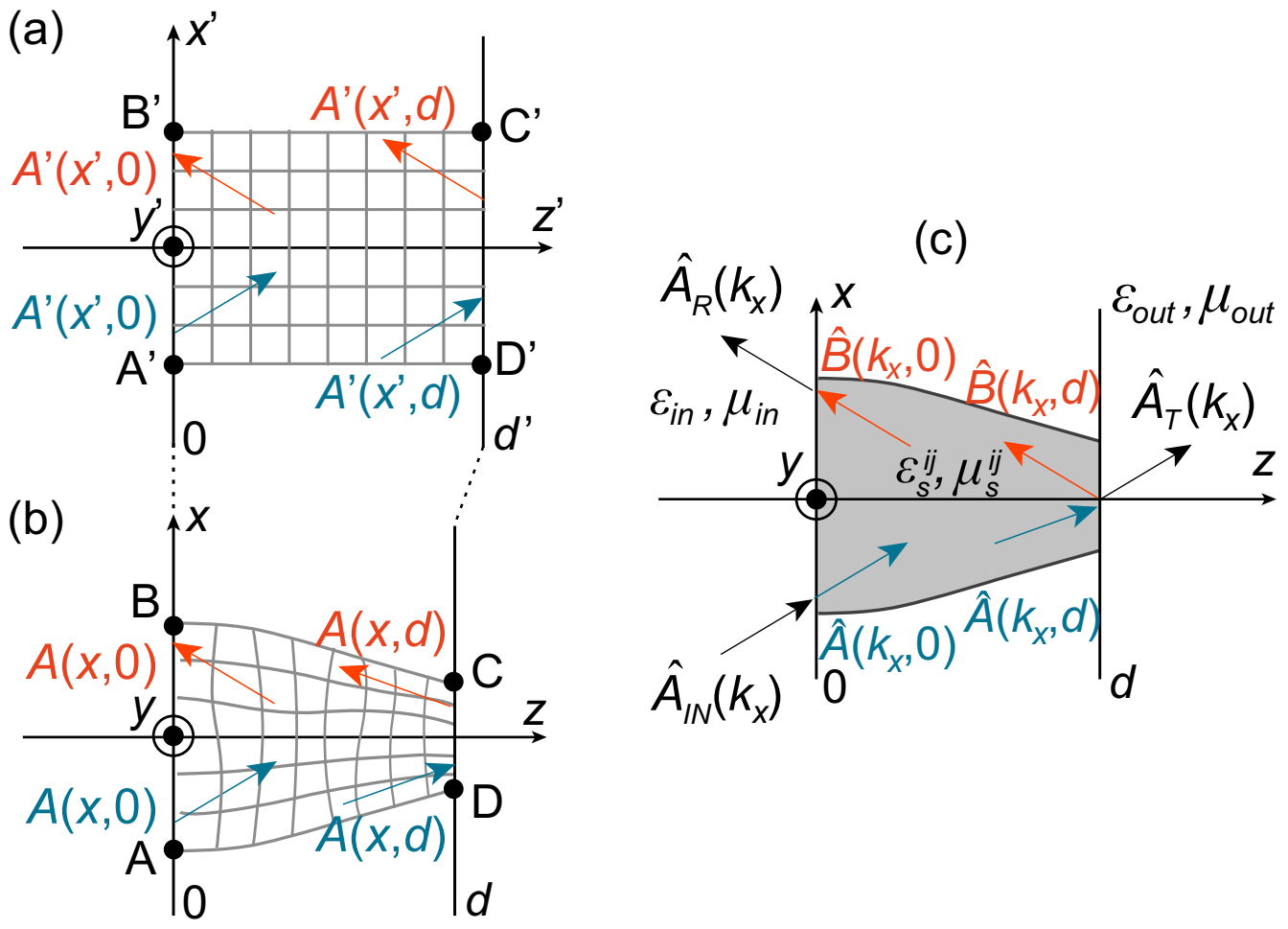

**Fig. 1** (a) Virtual space (b) Physical space (c) Squeezer and propagating waves inside and outside it.

The squeezer is placed between isotropic media characterized by relative constitutive parameters $\varepsilon_{in} = \mu_{in} = 1$ and $\varepsilon_{out}$, $\mu_{out}$. The field **A** is polarized along the $y$-direction (**A** = $A_y$**y**), with **A** being the electric (magnetic) field **E** (**H**) for TE (TM) waves. We denote the Fourier transform of $A_y(x,z)$ in the $x$-variable as $\hat{A}_y(k_x,z)$, where $k_x$ is the transverse component of the wavevector **k**. Our aim is to obtain the relation between incident, reflected and transmitted waves $\hat{A}_{IN}(k_x)$, $\hat{A}_R(k_x)$ and $\hat{A}_T(k_x)$ [see Fig. 1(c)]. A general transformation is given by $x = x(x',z')$, $y = y'$ and $z = z(x',z')$. The transformation can be whatever, except at the boundaries. In particular, we want the identity transformation at $z' = 0$ in order to avoid reflections at the input boundary, *i.e.*, $x(x',0) = x'$ and $z(x',0) = 0$. At $z' = d'$, we want the transformation to satisfy $\partial x/\partial x' = 1/F$ ($F$ is the compression factor along the $x$-direction) and $z(x',d') = d$. The first condition implies that the compression is uniform at the output interface. The second ensures that the output boundary is flat. With these simplifications, it can be shown that (derivation details are given in the appendix):

$$\begin{aligned} \hat{A}_R(k_x) &= \hat{A}_{IN}(k_x)R(k_xF)M^2(k_x) \\ \hat{A}_T(k_x) &= \frac{1}{F}\hat{A}_{IN}(k_x/F)T(k_x)M(k_x/F) \end{aligned}, \tag{1}$$

with $M(k_x) = \exp\left(i\sqrt{k^2-k_x^2}\,d'\right)$. For TM polarization:

$$R(k_x) = \frac{\varepsilon_{out}\sqrt{k_0^2F^2-k_x^2}-\sqrt{k_0^2\varepsilon_{out}\mu_{out}-k_x^2}}{\varepsilon_{out}\sqrt{k_0^2F^2-k_x^2}+\sqrt{k_0^2\varepsilon_{out}\mu_{out}-k_x^2}}. \tag{2}$$

$T(k_x)$ can be obtained from the relation $T(k_x) = 1 + R(k_x)$. The coefficients $R_2(k_x)$ and $T_2(k_x)$ for light impinging from the right at $z = d$ are given by $R_2(k_x) = -R(k_x)$ and $T_2(k_x) = 1 + R_2(k_x)$. To choose a specific transformation, we gave priority to those requiring only dielectric media, as the main difficulty at optical frequencies lies on achieving magnetic materials. Since we are dealing with a two-dimensional problem, only some components of the constitutive parameters need to be implemented. For TE polarization the tensor components that affect the fields are $\varepsilon^{yy}$, $\mu^{tt}$ and $\mu^{ll}$, being $t$ and $l$ the in-plane tensor principal directions. In this case, the problem can be solved by using quasi-conformal mappings, which give rise to $\mu^{tt} \approx \mu^{ll} \approx 1$ [18,20]. In the geometrical optics regime, such a medium should work for both polarizations. Nonetheless, an exact realization for TM waves involves the implementation of $\mu^{yy}$, $\varepsilon^{tt}$ and $\varepsilon^{ll}$. It can be shown that $\mu^{yy} = 1$ if we use a transformation of the form $x = x'f_1(z')$, $z = f_2(z')$, with $f_1(z') = (df_2(z')/dz')^{-1}$. Here we will employ the functions $f_1(z') = 1/(1+Cz')$ and $f_2(z') = z'(1+Cz'/2)$. This way, the exact realization of the squeezer only requires an anisotropic dielectric. For a squeezer length $d$ as small as 3 μm, only moderate refractive indices, approximately ranging from 0.5 to 2, are required. These values can be relaxed by using a lower compression factor or longer lengths. Moreover, $f_1(z')$ and $f_2(z')$ could be optimized to further adjust this range. The implementation of anisotropic dielectrics is feasible with the use of multilayer structures [9,19]. Remarkably, an exact dielectric realization of a compressing device working simultaneously for both polarizations could be achieved, since we only need to implement $\varepsilon^{tt}$ and $\varepsilon^{ll}$ for TM polarization and $\varepsilon^{yy}$ for TE polarization.

In order to verify Eqs. (1-2), a squeezer with $F$ = 2 embedded in air ($\varepsilon_{out}$ = $\mu_{out}$ = 1) was simulated with COMSOL Multiphysics for TM polarization (in this case $\hat{A}$ has to be replaced by $\hat{H}$). For the sake of computational and representation convenience, we rearrange Eqs. (1) for this specific case so that they read $R(2k_x) = \hat{H}_R(k_x)/\left(\hat{H}_{IN}(k_x)M^2(k_x)\right)$ and $T(2k_x)/2 = \hat{H}_T(2k_x)/\left(\hat{H}_{IN}(k_x)M(k_x)\right)$. Note that in the last equation we have scaled the independent variable by a factor of 2, which does not affect the meaning of the equation. Now we substitute in these equations the simulated values of $\hat{H}_{IN}(k_x)$, $\hat{H}_R(k_x)$ and $\hat{H}_T(k_x)$, as well as the theoretical values of $M(k_x)$, $R(k_x)$ and $T(k_x)$ as defined above. Finally, we depict in Fig. 2 the left and right hand sides of these equations for the case in which light impinges from the left. Figs. 2(a) and 2(b) show the absolute value of each member of the first and second equations, respectively (note that we have ignored $M(k_x)$ in this case, as this term only contributes to the phase response). In a similar way, Figs. 2(c) and 2(d) show the phase of each member of the first and second equations, respectively. The excellent agreement found between both sides of each equation confirms the validity of Eqs. (1-2). Similar results were obtained for light impinging from the right (not shown).

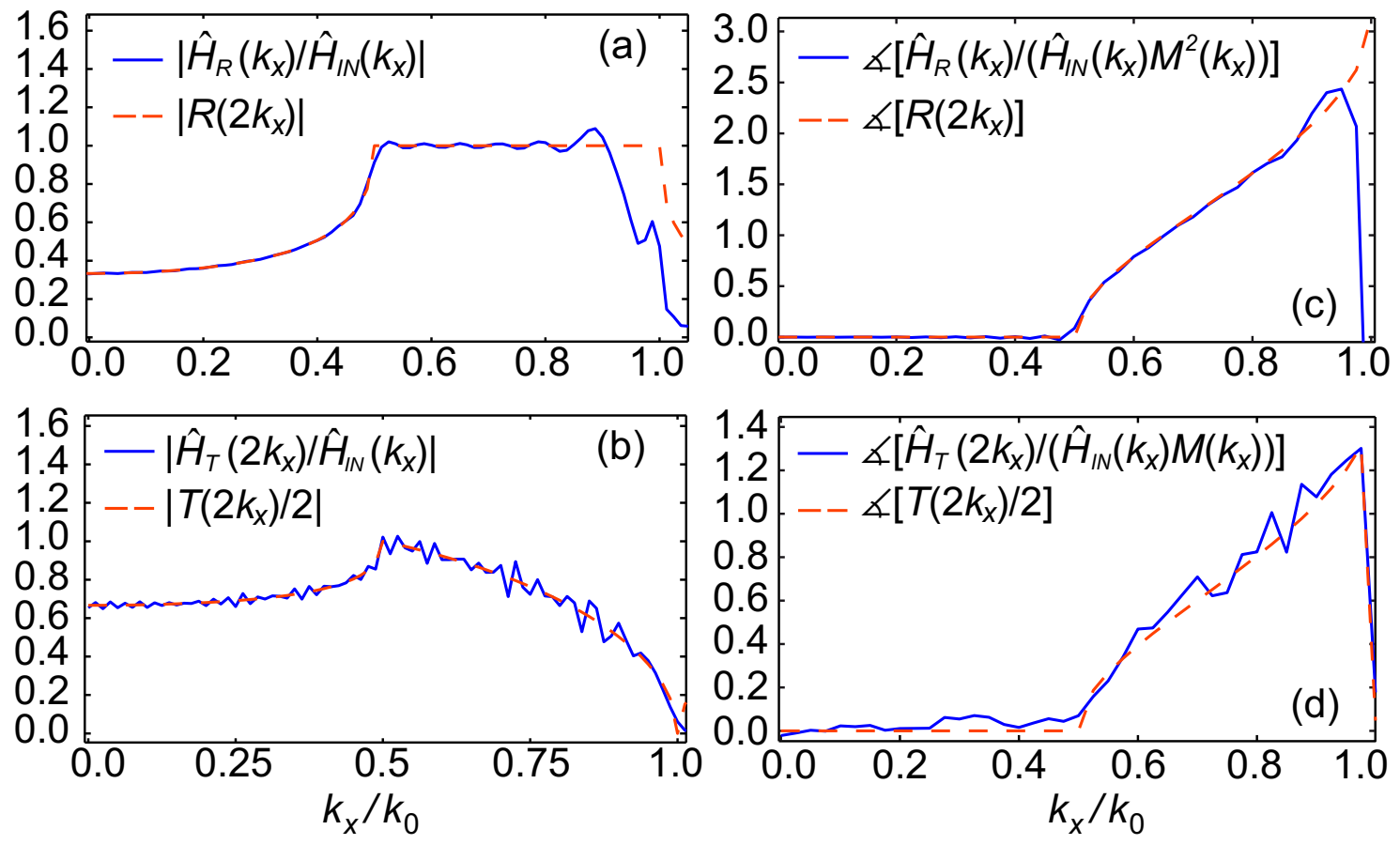

**Fig. 2** Numerical verification of Eqs. (1) and (2) for the case corresponding to a squeezer with $F = 2$ embedded in air and TM polarization. (a) Absolute value of the left and right hand side of equation $R(2k_x) = \hat{H}_R(k_x)/\left(\hat{H}_{IN}(k_x)M^2(k_x)\right)$ (b) Absolute value of the left and right hand side of equation $T(2k_x)/2 = \hat{H}_T(2k_x)/\left(\hat{H}_{IN}(k_x)M(k_x)\right)$ (c) Phase (represented by the symbol $\measuredangle$) of the left and right hand side of equation $R(2k_x) = \hat{H}_R(k_x)/\left(\hat{H}_{IN}(k_x)M^2(k_x)\right)$ (d) Phase of the left and right hand side of equation $T(2k_x)/2 = \hat{H}_T(2k_x)/\left(\hat{H}_{IN}(k_x)M(k_x)\right)$.

## Excitation of SPPs with a squeezing flat slab

As shown by Eq. (1), in addition to changing the amplitude of incident waves, the squeezer performs an expansion in $k_x$ by a constant factor $F$, whilst frequency is conserved. This change in the DR suggests that the squeezer could be used to excite SPPs. To illustrate this idea, we study the situation depicted in Fig. 3.

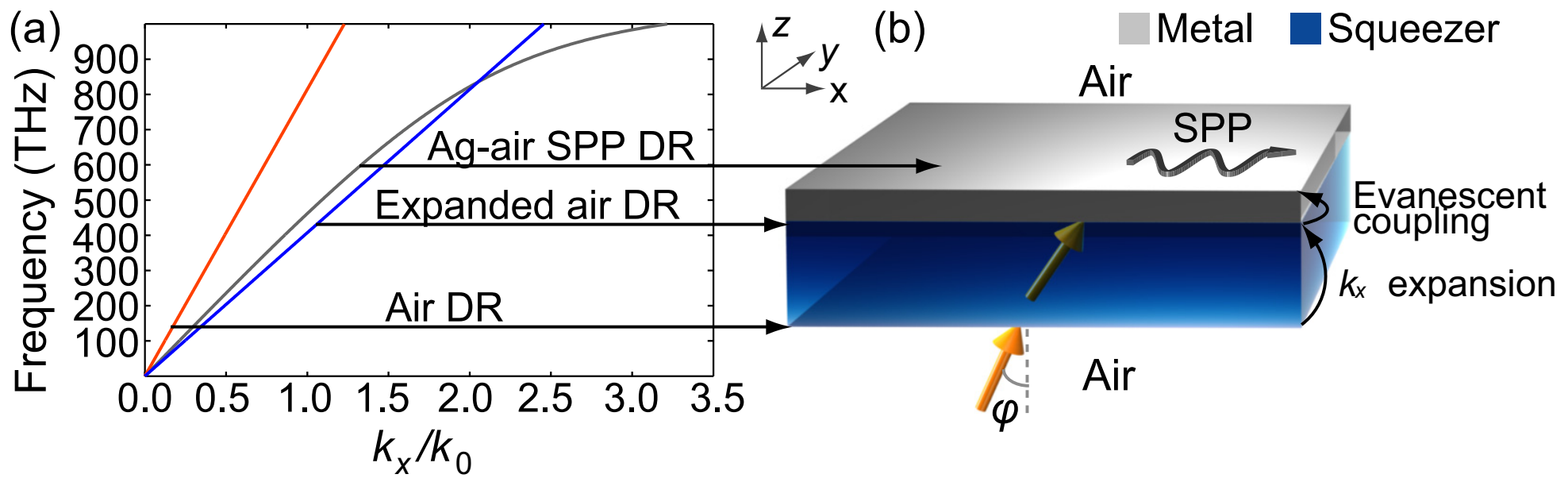

**Fig. 3** (a) Air light DR (orange), expanded air DR (blue), and metal-air SPP DR (grey) (b) SPPs excitation configuration.

A thin metal film (thickness $t$) is placed next to the output interface of a squeezer. The whole system is surrounded by air. Light impinges onto the squeezer at an angle $\varphi$ with $k_x = k_0\sin(\varphi)$, where $k_0 = 2\pi/\lambda$ and $\lambda$ is the free-space wavelength. The curve associated to its DR and that of an SPP propagating along the metal-air interface never cross and direct coupling is not possible. However, after passing through the squeezer, the DR of free-space light is scaled in $k_x$ by a factor $F$ and intersects the SPP DR curve. Thus, light exiting the squeezer can tunnel through the metal and excite the SPP along the metal-air surface via evanescent coupling. Since the squeezer expands $k_x$ by $F$, this is analogous to using a prism with refractive index $F$ in the Kretschmann configuration. As an example, we consider a typical situation with $\lambda$ = 632.8 nm (HeNe laser), $F$ = 1.766 (Sapphire refractive index), and $t$ = 45 nm [13]. As for the metal, we use Ag, whose permittivity was found to be $\varepsilon_m = -15.98 + 0.72i$ at this wavelength in [13]. We can deduce the value of $k_x$ for which the SPP is excited at $\lambda$ from the reflection coefficient $\Gamma(k_x)$ of the system squeezer-metal-air. Since there are no reflections at the input, we only need to take into account the squeezer-metal and metal-air interfaces. By using Eq. (2) and Fresnel equations we calculated $\Gamma(k_x)$ (see Fig. 5). The SPP excitation is stronger at the $\Gamma(k_x)$ minimum, at $k_x = 1.029\cdot10^7$ rad/m. This is also the value of $k_x$ at the squeezer output. According to Eq. (1), this corresponds to $k_x = 1.029\cdot10^7/1.766 = 5.83\cdot10^6$ rad/m for incident light ($\varphi \approx 36°$). Finally, we simulate this example. A Gaussian beam impinging onto the squeezer at $\varphi = 36°$ is used as the source. SPPs only exist for TM polarization, for which we check our device. The simulated power flow is rendered in Fig. 4(a). The beam smoothly enters the squeezer, reaches the metal layer and tunnels through it, launching the SPP.

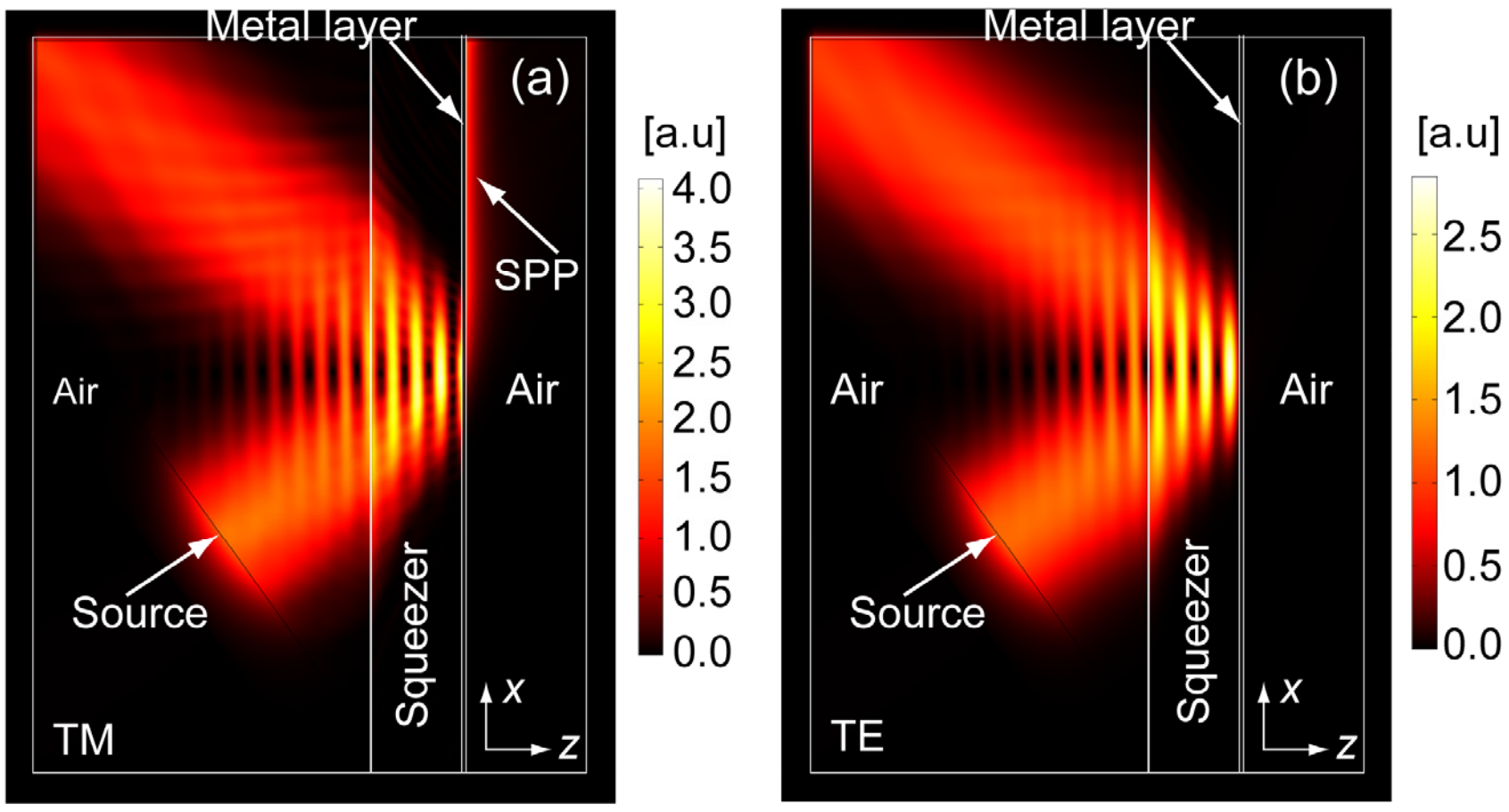

**Fig. 4** Simulated SPP excitation ($d$ = 1.5 μm). Absolute value of the power flow for (a) TM and (b) TE polarization.

As a verification test, we repeated the simulations for TE polarization [Fig. 4(b)]. As expected, no SPP is excited in this case. It is worth noting that, unlike in prism couplers, no Fabry-Perot interference exists within the squeezer because no reflections occur at its input interface. As a consequence, the power transferred to the SPP does not depend on the size of the squeezer. In this sense, the proposed excitation configuration mimics a dielectric-metal-air system with the source embedded inside the dielectric [13], but with the advantage that the source can be placed outside the system without the need of giving the dielectric a prism shape.

## Broadening the angular bandwidth

According to Fig. 3, SPPs can be excited only at the angle whose associated $k_x$ matches that of the SPP. In a more realistic picture, the SPP resonance has a certain width, as seen in Fig. 5, where $\Gamma(k_x)$ is depicted (the coefficient for a system dielectric-metal-air turns out to be the same if the dielectric refractive index is $F$). The strength of the excitation is higher at the minimum of $\Gamma(k_x)$, but waves with $k_x$ around the optimal one also excite the SPP to a lesser extent. Moreover, in many cases the incident wave is not a plane wave and its power spectral density (PSD) spreads over a finite region in $k_x$ (for instance, when using a laser or the output of an optical fiber). This allows the excitation of SPPs at a set of angles determined by the source distribution in the angular spectrum. To analyze the angular bandwidth of the proposed device and compare it with that of a prism coupler, we consider the case in which the source is a Gaussian beam with the same excitation configuration as in Fig. 4. The beam axis always crosses the point $x = 0$, $z = d$. In order to avoid Fabry-Perot resonances, we model the prism coupler as a semi-infinite dielectric medium that extends over the interval $z \in (-\infty, d)$, with the source embedded within the dielectric (configuration of Fig. 4, but replacing the squeezer and left air region by the dielectric). To be comparable with the squeezer, the dielectric refractive index is taken to be equal to the compression factor $F$ (see discussion above). In Fig. 5(a) we show the profile of the field that impinges onto the metal film in each case (at $z = d$), which we call $A(x,d)$ to follow the notation of Fig. 1.

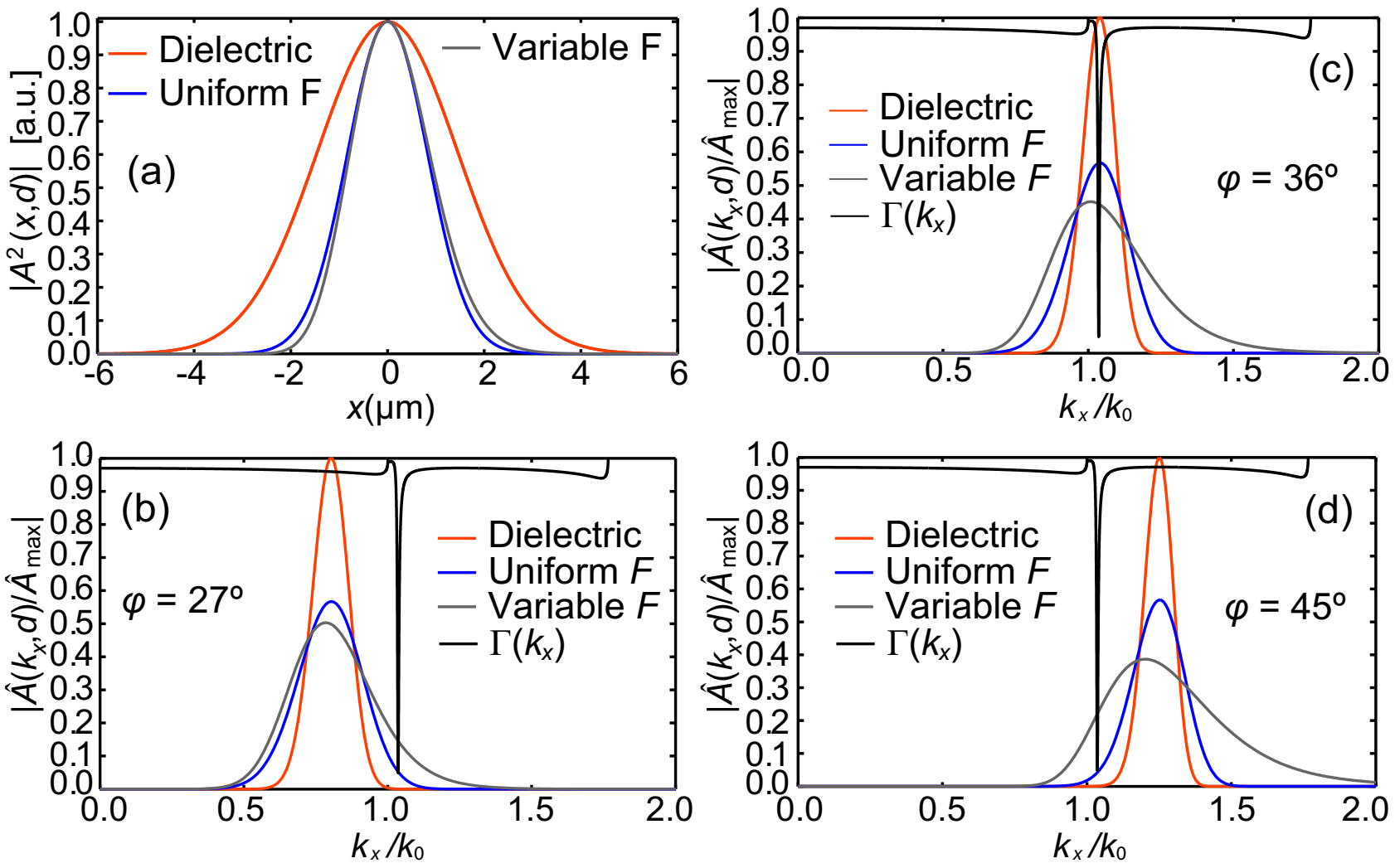

**Fig. 5** (a) Field $A(x,d)$ impinging onto the metal film after exiting the dielectric coupler (orange), uniform squeezer (blue) and variable squeezer (grey) (b-d) Corresponding $\hat{A}(k_x,d)$ normalized to its maximum value $\hat{A}_{\max}$ (reached by the dielectric coupler) at different angles of incidence $\varphi$. $\Gamma(k_x)$ is shown in black.

The beam exiting the squeezer is a compressed version of that exiting the dielectric. Therefore, its associated spectral distribution $\hat{A}(k_x,d)$ is wider than that of the dielectric. This results in a broader angular bandwidth, as we can infer from Fig. 5. When $\varphi$ is the optimal one [Fig. 5(c)], both spectral distributions are centered at the minimum of $\Gamma(k_x)$, providing the maximum excitation intensity. Naturally, the excitation strength is higher in the case of the prism, since the PSD of the wave exiting the squeezer is lower due to the spreading in $k_x$ that it performs. However, when we decrease $\varphi$ [Fig. 5(b)], both spectral distributions shift to the left. Now, the PSD overlapping the dip in $\Gamma(k_x)$ is very low in the prism case. Due to its broader spectral extension, the PSD is much higher in the squeezer case, and so is the excitation strength. The situation is similar at larger values of $\varphi$ [Fig. 5(d)]. It is worth mentioning that the coupling efficiency can be significantly lower in a real prism coupler than in the considered ideal case, depending on the reflections that take place at the input interface. Moreover, due to these reflections, a secondary SPP can be excited in the undesired direction. To quantify the angular bandwidth of dielectric and squeezer couplers, we performed a series of simulations at different angles. The source field distribution is the same for both couplers, except for a multiplicative constant, to ensure that the power radiated by the source is the same in all cases. In Fig. 6(a) we depict the electric field amplitude of the excited SPP at the metal surface as a function of $\varphi$. For the dielectric coupler, the half-power angular bandwidth (BW) was found to be 4º, while in the case of the squeezer, a value of 7.1º was obtained. These results are explained by the previous discussion and the information in Fig. 5.

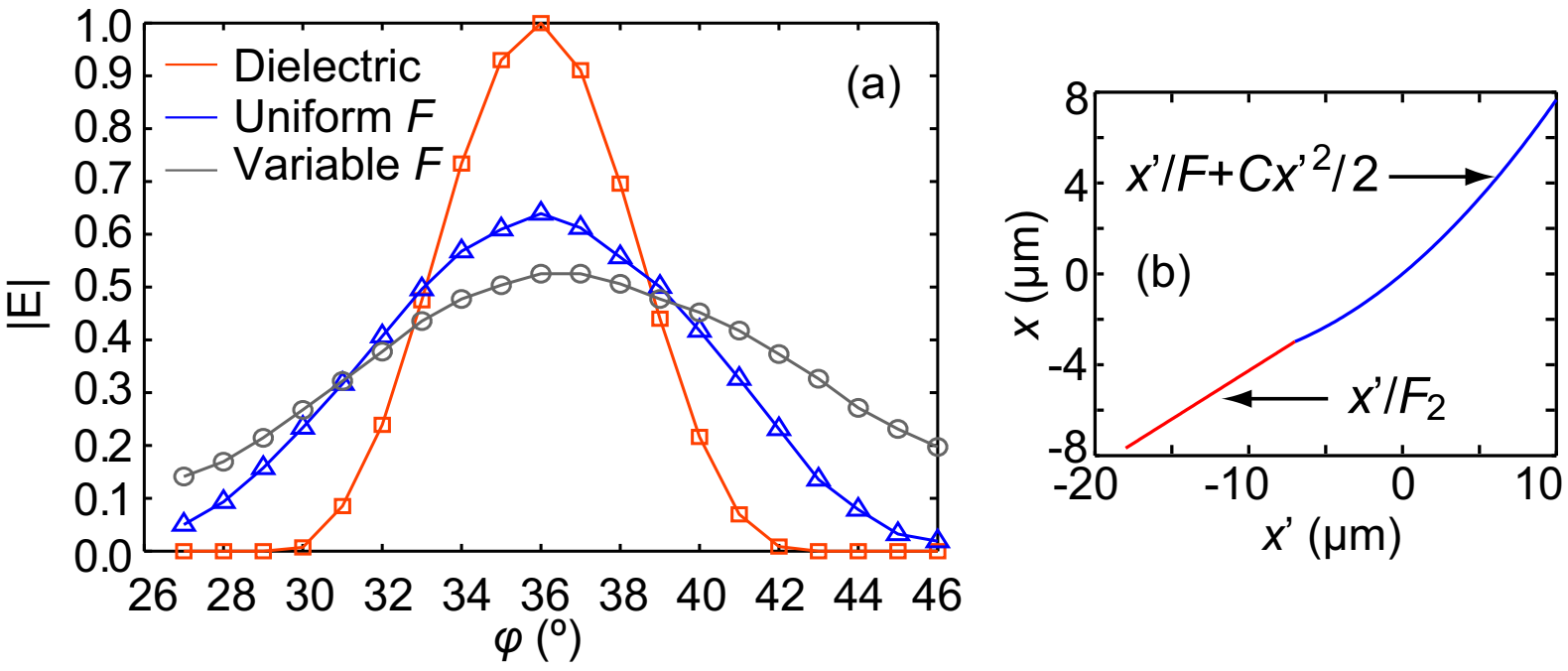

**Fig. 6** (a) Electric field amplitude of the excited SPP at the metal surface as a function of $\varphi$. Orange: dielectric coupler. Blue: uniform squeezer. Grey: variable squeezer (b) Transformation used for the variable squeezer at $z = d$.

We have seen that the squeezer coupler presents a higher BW because of the compression in $x$ (expansion in $k_x$) that it performs. This suggests that we could control the device BW by engineering the way in which the fields are transformed. For instance, we could further increase the BW by allowing a compression factor that varies with the $x$-coordinate at the squeezer output. Intuitively, a variable compression introduces more spatial frequencies and the transformed beam should have a broader associated PSD. As an example, consider the transformation $x = x'(1-z/d) + zg(x')/d$, $y = y'$, and $z = z'$, with $g(x') = x'/F + Cx'^2/2$ for $x' \geq x_0$ and $g(x') = x'/F_2$ for $x' < x_0$, $C$ being a constant. At $z = 0$ we have the identity transformation. At $z = d$ and for $x' \geq x_0$ the compression varies linearly with $x'$, being equal to $F$ at $x = 0$. In order to keep this variable compression factor within a reasonable range of values, our transformation only performs a uniform scaling for $x' < x_0$. $F_2$ is chosen so that the transformation is continuous at $x_0$. In Fig. 5(a) we render the appearance of the previous Gaussian beam after undergoing this transformation with $F = 1.766$, $F_2 = 2.346$, $C = 4\cdot10^5$ $\mathrm{m}^{-1}$ and $x_0 = -7$ μm. The transformation at $z = d$ is shown in Fig. 6(b). Owing to the non-uniform compression, the transformed beam has no symmetry with respect to the origin. As a consequence, $\hat{A}(k_x, d)$ is also non-symmetric and extends over a broader spectral range. At the optimal coupling angle, the PSD of the wave exiting the variable squeezer is slightly lower than that of the uniform one [Fig. 5c]. However, at angles away from the optimal one, the PSD is significantly larger for the variable squeezer [Fig. 5(b,d)], which implies a widening of the angular bandwidth. Although the results in Fig. 5 only provide a qualitative description (the reflection coefficients of both squeezers are different and this affects the coupling efficiency) they capture the essential features of both devices. Simulations confirmed the behavior predicted for the variable squeezer (see Fig. 6), which reaches a BW of 10.1º (2.5 times that of the dielectric coupler). Note that the variable squeezer always broadens the angular bandwidth (even if the source is a plane wave). This technique not only allows us to broaden the BW, but also to achieve other coupler angular responses by using different mappings that properly transform the source PSD. Although we cannot use the previous dielectric implementation for the variable squeezer [the transformation is not of the form $x = x'f_1(z')$], a dielectric realization could still be possible, for instance, by allowing a curved output boundary.

## Conclusions

In summary, we have presented an alternative way of exciting SPPs with the help of TO. To this end, we exploit the modification in the DR introduced by a squeezing device. The squeezer plays the role of a prism coupler but is more suitable for integration because of its flat geometry and small dimensions. Moreover, it does not introduce reflections at its input interface, preventing the appearance of Fabry-Perot resonances and the excitation of SPPs in the undesired direction. The implementation of the squeezer can be achieved with dielectric anisotropic materials. Finally, we have shown that the coupler angular bandwidth can be broadened by using a variable compression. As an example, we have designed a device with a BW 2.5 times higher than that of a conventional dielectric coupler. This technique could pave the way for engineering SPP couplers with tailor-made angular responses.

**Acknowledgements** Financial support by Spanish Ministerio de Ciencia e Innovación (contracts CSD2008-00066 and TEC2008-06871-C02, and FPU grant) is gratefully acknowledged.

## Appendix

A detailed derivation of Eq. (1) is provided next. The fields propagating in physical space can be expressed as a function of those in virtual space. In component notation:

$$A_i(x,z) = \Lambda_i^{i'} A_{i'}\left(x'(x,z), z'(x,z)\right), \tag{3}$$

with $\Lambda_i^{i'} = \partial x_i / \partial x_{i'}$, $i$ = 1, 2, 3, and $x_1 = x$, $x_2 = y$, $x_3 = z$, with analogous expressions for primed coordinates. Since we have the identity transformation at $z' = 0$, then $x(x',0) = x'$ and $z(x',0) = 0$. Therefore, using Eq. (3) and taking into account that the only non-vanishing component of **A** is $A_2 = A_y$:

$$A_y(x,0) = \Lambda_2^{i'} A_{i'}\left(x'(x,0), z'(x,0)\right) = A_{y'}(x,0). \tag{4}$$

At $z' = d'$ we also have the restrictions $\partial x/\partial x' = 1/F$ and $z(x',d') = d$. The first condition implies that the compression is uniform at the device output interface, leading to $x = x'f(z')$ at $z' = d'$, *i.e.*, $x(x',d') = x'/F$, if there is no translation in $x$ at this point. With these simplifications, we can write:

$$A_y\left(x,d\right)=\Lambda_2^{i'}A_{i'}\left(x'\left(x,d\right),z'\left(d\right)\right)=A_{y'}\left(xF,d'\right). \tag{5}$$

Our goal is to relate the fields transmitted and reflected by the squeezer with the corresponding incident field as a function of $k_x$. For this purpose, we will first calculate in Fourier space the relation between the fields at $z = 0$ and $z = d$ inside the device. In virtual space (empty flat space), the relation between the fields at $z' = 0$ and $z' = d'$ is well known. Specifically, $\hat{A}_{y'}\left(k_{x'},d'\right)=\hat{A}_{y'}\left(k_{x'},0\right)M(k_{x'})$ when propagation is towards increasing values of $z$ [blue arrows in Fig. 1(a)] and $\hat{A}_{y'}\left(k_{x'},0\right)=\hat{A}_{y'}\left(k_{x'},d'\right)M(k_{x'})$ when propagation is towards decreasing values of $z$ [orange arrows in Fig. 1(a)]. $M(k_x)$ is the transfer function of a free space slab of thickness $d'$, *i.e.*, $M(k_x) = \exp\left(i\sqrt{k^2-k_x^2}d'\right)$. With the aid of Eqs. (4-5), we find for waves propagating towards increasing values of $z$ [blue arrows in Fig. 1(b)]:

$$\begin{aligned}\hat{A}_y\left(k_x,d\right)&=\int_{-\infty}^{\infty}A_y\left(x,d\right)e^{-ik_xx}dx=\int_{-\infty}^{\infty}A_{y'}\left(xF,d'\right)e^{-ik_xx}dx=\\&=\frac{1}{F}\hat{A}_{y'}\left(k_x/F,d'\right)=\frac{1}{F}\hat{A}_y\left(k_x/F,0\right)M\left(k_x/F\right)\end{aligned}. \tag{6}$$

For waves propagating towards decreasing values of $z$ [orange arrows in Fig. 1(a-b)]:

$$\left.\begin{aligned}\hat{A}_y\left(k_x,0\right)&=\hat{A}_{y'}\left(k_x,d'\right)M\left(k_x\right)\\ \hat{A}_y\left(k_x,d\right)&=\frac{1}{F}\hat{A}_{y'}\left(k_x/F,d'\right)\end{aligned}\right\}\Rightarrow\hat{A}_y\left(k_x,0\right)=F\hat{A}_y\left(k_xF,d\right)M\left(k_x\right). \tag{7}$$

Now we consider the squeezer as a block that returns two outputs (reflected and transmitted waves $A_R$ and $A_T$) under an input (incident wave $A_{IN}$). The relation between $\hat{A}_{IN}\left(k_x\right)$, $\hat{A}_R\left(k_x\right)$ and $\hat{A}_T\left(k_x\right)$ can be obtained with the help of Fig. 1(c). As mentioned in the main text, the squeezer is placed between isotropic media characterized by relative constitutive parameters $\varepsilon_{\text{in}}$, $\mu_{\text{in}}$ (input medium) and $\varepsilon_{\text{out}}$, $\mu_{\text{out}}$ (output medium) and we consider the input medium to be the same as in virtual space ($\varepsilon_{\text{in}} = \mu_{\text{in}} = 1$). This way, the transformation is continuous at $z = 0$ and there occur no reflections at this interface. Thus, $\hat{A}_{IN}\left(k_x\right)=\hat{A}\left(k_x,0\right)$ and $\hat{A}_R\left(k_x\right)=\hat{B}\left(k_x,0\right)$. However, the transformation discontinuity at $z = d$ introduces reflections at the (squeezer)-(output medium) interface. The relation between a wave impinging from the left at $z = d$ and the transmitted and reflected waves can be characterized by a pair of reflection and transmission coefficients $R(k_x)$ and $T(k_x)$, in such a way that $\hat{A}_T\left(k_x\right)=\hat{A}\left(k_x,d\right)T(k_x)$ and $\hat{B}\left(k_x,d\right)=\hat{A}\left(k_x,d\right)R(k_x)$. Moreover, we know that $\hat{A}\left(k_x,d\right)=F^{-1}\hat{A}\left(k_x/F,0\right)M\left(k_x/F\right)$ from Eq. (6), and that $\hat{B}\left(k_x,0\right)=F\hat{B}\left(k_xF,d\right)M\left(k_x\right)$ from Eq. (7). According to these relations:

$$\hat{A}_R(k_x) = \hat{A}_{IN}(k_x) R(k_x F) M^2(k_x)$$
$$\hat{A}_T(k_x) = \frac{1}{F} \hat{A}_{IN}(k_x/F) T(k_x) M(k_x/F) \quad (8)$$